\documentclass{iopart}
\usepackage{iopams}
\usepackage{setstack}
\usepackage{graphicx}
\usepackage{bm}
\usepackage[usenames]{color}
\bibstyle{apsrev.bib}

\newcommand{\be}{\begin{equation}}
\newcommand{\ee}{\end{equation}}
\newcommand{\beqn}{\begin{eqnarray}}
\newcommand{\eeqn}{\end{eqnarray}}

\begin{document}

\title{Nonequilibrium quench dynamics in quantum quasicrystals}

\author{Ferenc Igl\'oi$^{1,2}$, Gerg\H o Ro\'osz$^{2,1}$, Yu-Cheng Lin$^3$}
\address{$^1$ Wigner Research Centre, Institute for Solid State Physics and Optics,
H-1525 Budapest, P.O.Box 49, Hungary}
\address{$^2$ Institute of Theoretical Physics,
Szeged University, H-6720 Szeged, Hungary}
\address{$^3$ Graduate Institute of Applied Physics,
National Chengchi University, Taipei, Taiwan}
\ead{igloi.ferenc@wigner.mta.hu}
\ead{gergo\_roosz@titan.physx.u-szeged.hu}
\ead{yc.lin@nccu.edu.tw}


\begin{abstract}
We study the nonequilibrium dynamics of a quasiperiodic quantum Ising chain after a sudden
change in the strength of the transverse field at zero temperature. In particular we consider
the dynamics of the entanglement entropy and the relaxation of the magnetization.
The entanglement entropy increases with time as a power-law, and
the magnetization is found to exhibit stretched-exponential relaxation.
These behaviors are explained in terms of anomalously diffusing quasiparticles, which
are studied in a wave packet approach. The nonequilibrium magnetization is shown to have
a dynamical phase transition.
\end{abstract}

\maketitle
\section{Introduction}
\label{sec:intr}
Recent experimental progress in ultracold atomic gases in optical
lattices \cite{Greiner_02,Paredes_04,Kinoshita_04,Kinoshita_06,Lamacraf_06,Sadler_06,Hofferberth_07,Trotzky_12,Cheneau_12,Gring_11}
has opened up fascinating new perspectives on research in the field of isolated
quantum systems, both in equilibrium and out of equilibrium.  In experiments
the form of atomic interactions can be suddenly changed by tuning an applied
magnetic field near a Feshbach resonance, which is known as a global quantum
quench. On the theoretical side, one is interested in the time-evolution of
different observables, such as the order parameter or some correlation function,
after a quench. Fundamental questions concerning quantum quenches include
(i) the functional form of the relaxation process in early times, 
and (ii) the properties of the stationary state of the system 
after a sufficiently long time.  

Many results for quantum quenches have been obtained for \textit{homogeneous}
systems \cite{Polkovnikov_11,Rigol_07,Calabrese_06,Calabrese_07,Cazalilla_06,Manmana_07,Cramer_08,Barthel_08,Kollar_08,Sotiriadis_09,Roux_09,
Sotiriadis_11,Kollath_07,Banuls_11,Gogolin_11,Rigol_11,Caneva_11,Cazalilla_11,Rigol_12,Santos_11,Grisins_11,Canovi_11};
for example, the relaxation of correlation functions in space and in time is
generally in an exponential form, which defines a quench-dependent correlation
length and a relaxation (or decoherence) time. Many basic features of
the relaxation process can be successfully explained by a quasiparticle
picture \cite{Calabrese_05,Calabrese_07,Rieger_11}: after a global quench quasiparticles are
created homogeneously in the sample and  move ballistically with momentum
dependent velocities.  The behavior of observables in the stationary state is
generally different in integrable and in non-integrable systems. For
non-integrable models, thermalization is
expected \cite{Rigol_07,Calabrese_06,Calabrese_07,Cazalilla_06,Manmana_07,
Cramer_08,Barthel_08,Kollar_08,Sotiriadis_09,Roux_09,Sotiriadis_11} and the
distribution of an observable is given by a thermal Gibbs ensemble;
however, in some specific examples this issue has turned out to be more
complex \cite{Kollath_07,Banuls_11,Gogolin_11,Grisins_11}.
By contrast, it was conjectured that  
stationary state averages for integrable models are described by a generalized 
Gibbs ensemble \cite{Rigol_07}, in which each integral of motion is separately
associated with an effective temperature.   

Concerning quantum quenches in inhomogeneous systems, there have been only a
few studies in specific cases; for example, entanglement entropy dynamics in
random quantum chains \cite{dyn06,Igloi_12,Levine_12} and in models of
many-body localization \cite{Pollman_12,Vosk_12}.  In some of these cases the
eigenstates are localized, which prevents the system from reaching a thermal
stationary state. 

A special type of inhomogeneity, interpolating between homogeneous and
disordered systems, is a quasicrystal \cite{Shechtman,dubois} or an aperiodic
tiling \cite{Penrose}. Quasicrystals are known to have anomalous transport
properties \cite{Stadnik,Mayou}, which is due to the fact that in these systems
the long-time motion of electrons is not ballistic, but 
an anomalous diffusion described by a power law. One may expect that the quasiparticles 
created during the quench have a similar dynamical behavior,
which in turn affects the relaxation properties of quasicrystals.

Quasicrystals of ultracold atomic gases have been experimentally realized in
optical lattices by superimposing two periodic optical waves with different
incommensurate wavelengths. An optical lattice produced in this way realizes a
Harper's quasiperiodic potential \cite{Harper,Aubry}, for which the eigenstates
are known to be either extended or localized depending on the strength of the
potential. Different phases of Bose-Hubbard model with such a potential have
been experimentally investigated \cite{Roati,Deissler}. There have also been
theoretical studies concerning the relaxation process in the Harper
potential \cite{Modugno_09,Gramsch}.

In this paper we consider the nonequilibrium quench dynamics of the quantum
Ising chain in one-dimensional quasicrystals.  The quantum Ising chain in its
homogeneous version is perhaps the most studied model for nonequilibrium
relaxation \cite{Barouch_70,Igloi_00,Sengupta_04,Fagotti_08,Silva_08,Rossini_09,
Campos_Venuti_10,Igloi_11,Foini_11,Divakaran_11,Rieger_11,Calabrese_11,Schuricht_12,
Calabrese_12,Calabrese_12b,blass,Essler_12}. Our study extends previous
investigations in several respects and seeks to obtain new insights into quench
dynamics in inhomogeneous systems.  We focus on the Fibonacci lattice, for which
many equilibrium properties of the quantum Ising model are
known \cite{igloi88,turban94,igloi96b,it2,hgb,hermisson,IJZ07}; to our knowledge
this is the first study of quantum quenches in such a lattice.  Using
free-fermionic techniques \cite{lieb61}, we numerically calculate the
time-dependence of the entanglement entropy as well as the relaxation of the
local magnetization for large lattices.  The numerical results are interpreted
by a modified quasiparticle picture, in which the quasiparticles are represented by wave packets; we
also obtain diffusive properties of the wave packets.

The structure of the paper is as follows. The quasiperiodic quantum Ising model
and its equilibrium properties are described in section~\ref{sec:model}. The global
quench process and some known results for homogeneous and random chains are
presented in section~\ref{sec:nonequilibrium}.  Our numerical results for the
quasiperiodic chain are presented and interpreted in section~\ref{sec:quasi}. This
paper is concluded with a discussion; some details of the free-fermionic calculation of
the local magnetization are presented in the appendix.

\section{The Model and its equilibrium properties}
\label{sec:model}
We consider the quantum (or transverse) Ising model defined by the Hamiltonian:
\be
{\cal H}=-\frac{1}{2}\left[ \sum_i {J}_i \sigma_i^x \sigma_{i+1}^x + h \sum_i  \sigma_i^z \right] \;,
\label{hamilton}
\ee
where $\sigma_i^{x}$ and $\sigma_i^{z}$ are Pauli matrices at site $i$. The interactions, $J_i$, are
generally site dependent, which are parameterized as:
\be
J_i=J r^{f_i}\;,
\label{J_i}
\ee
where $r>0$ is the amplitude of the inhomogeneity, and the integers $f_i$ are taken from a
quasiperiodic sequence.

Quasiperiodic lattices can be generated in different ways, such as by the cut-and-project method.
Here we use the following algebraic definition for a one-dimensional quasiperiodic sequence:
\be
f_i=1+\left[ \frac{i}{\omega}\right]-\left[ \frac{i+1}{\omega}\right] \;,
\label{f_i}
\ee
where $[x]$ denotes the integer part of $x$, and $\omega>1$ is an irrational number. The
Fibonacci sequence generated by the substitution rule: $0 \to 01$ and $1 \to 0$ starting
with $0$ corresponds to the formula in (\ref{f_i}) with the golden mean $\omega=(\sqrt{5}+1)/2$.
The parameter $J$ in (\ref{J_i}) is fixed with $J=r^{-\rho}$, where
\be
\rho=\lim_{L \to \infty} \frac{\sum_{i=1}^L f_i}{L}=1-\frac{1}{\omega} \;,
\label{rho}
\ee
is the fraction of units $1$ in the infinite sequence. Note that $r=1$ represents the homogeneous lattice.

The essential technique in the solution of ${\cal H}$ is the mapping to
spinless free fermions \cite{lieb61,pfeuty79}. First we express the spin operators
$\sigma_i^{x,y,z}$ in terms of fermion creation (annihilation) operators
$c_i^\dagger$ ($c_i$) by using the Jordan-Wigner
transformation \cite{JW}:  $c^\dagger_i=a_i^+\exp\left[\pi \imath \sum_{j}^{i-1}a_j^+a_j^-\right]$
and $c_i=\exp\left[\pi \imath
\sum_{j}^{i-1}a_j^+a_j^-\right]a_i^-$, where $a_j^{\pm}=(\sigma_j^x \pm \imath\sigma_j^y)/2$. 
Here and throughout the paper we denote the imaginary unit $\sqrt{-1}$ by $\imath$ to
avoid confusion with the integer index $i$.
The Ising Hamiltonian in (\ref{hamilton}) can then be written in a quadratic form in fermion operators.
\beqn
{\cal H}&=&
-\sum_{i=1}^{L}h \left( c^\dagger_i c_i-\frac{1}{2} \right) -
\frac{1}{2}\sum_{i=1}^{L-1} J_i(c^\dagger_i-c_i)(c^\dagger_{i+1}+c_{i+1})\cr
&+&\frac{1}{2} J_L(c^\dagger_L-c_L)(c^\dagger_{1}+c_{1})\exp(\imath\pi \mathcal{N}),
\label{ferm_I}
\eeqn
where $\mathcal{N}=\sum_{i=1}^L c_i^\dag c_i$ is the number of fermions.
The Hamiltonian (\ref{ferm_I}) can be diagonalized through a canonical transformation \cite{lieb61}, 
in which a new set of fermion operator $\eta_k$ is introduced by
\be
     \eta_k=\sum_{i=1}^L\left[ \frac{1}{2}\left(\Phi_k(i)+\Psi_k(i)\right)c_i+
   \frac{1}{2}\left(\Phi_k(i)-\Psi_k(i)\right)c_i^\dagger\right]
    \label{eta_ferm}
\ee
where the $\Phi_k(i)$ and $\Psi_k(i)$ are real, and normalized by
\be
     \sum_{k=1}^L \Phi_k(i)\Phi_k(j)=\sum_{k=1}^L\Psi_k(i)\Psi_k(j)=\delta_{ij}\;.
    \label{normalization}
\ee
We then obtain the diagonal form of ${\cal H}$:
\be
{\cal H}=\sum_{k=1}^L \epsilon_k \left( \eta_k^{\dag} \eta_k -\frac{1}{2}\right) \;,
\label{H_free}
\ee
in terms of the new fermion creation (annihilation) operators $\eta_k^\dag$ ($\eta_k$). 
The energies of free fermionic modes, $\epsilon_k$, and the components, $\Phi_k(i)$ and $\Psi_k(i)$, 
can be obtained from the solutions of the eigenvalue problem:
\begin{eqnarray}
\epsilon_k\Psi_k(i)&=&-h\Phi_k(i)-J_k\Phi_k(i+1)\; ,\nonumber\\
\epsilon_k\Phi_k(i)&=&-J_{k-1}\Psi_k(i-1)-h\Psi_k(i)\;.
\label{Phi_Psi}
\end{eqnarray}

The spectrum of free-fermionic excitations, $\epsilon_k$ in (\ref{H_free}),
plays a key role in equilibrium and non-equilibrium properties of the system.
In equilibrium and in the thermodynamic limit the model has a quantum critical
point at $h=h_c$, the properties of which are controlled by the low-energy
excitations. The value of $h_c$ is determined by the equation \cite{pfeuty79}
$\ln h_c=\overline{\ln J}$, where the overbar denotes an average over all
sites.  With the parameterization given above, the critical point is given by
$h_c=1$, independently of $r$.  The lowest gap, $\Delta E=\epsilon_1$, is zero
for $h<h_c$, and vanishes as $\Delta E \sim (h_c-h)^{\nu}$, as $h$
approaches $h_c$.  The singularity of the gap, measured by the gap-exponent
$\nu=1$, does not depend on $r$; the same is true for the singularity of the
specific heat: $C_v \sim \ln |h-h_c|$. Thus the transition belongs to the
Onsager universality class \cite{ONSAGER}, irrespectively of $r$.  This means
that the quasiperiodic modulation of the couplings represents an irrelevant
perturbation at the critical point of the homogeneous model \cite{luck93b}. For
$h<h_c$ the system is in the ordered phase, so that the local magnetization at
site $l$ is $m_l>0$. Upon approaching the critical point, the local
magnetization goes to zero following a power law: the bulk magnetization
$m_\mathrm{b}$ decays as $m_\mathrm{b}(h) \sim (h_c-h)^{1/8}$, which defines
the critical exponent $\beta_\mathrm{b}=1/8$, while the surface magnetization
$m_1$ vanishes as $m_1(h) \sim (h_c-h)^{\beta_\mathrm{s}}$ with
$\beta_\mathrm{s}=1/2$.  For $h>h_c$ the system is in the disordered phase and
the local magnetization vanishes in the thermodynamic limit.
 
While in equilibrium only the low-energy excitations are of importance,
the complete energy spectrum contributes to nonequilibrium properties,
which are investigated in this paper. 

\section{Nonequilibrium properties of homogeneous and random chains}
\label{sec:nonequilibrium}

We consider a quench process in which at time $t=0$ the strength of the
transverse field is changed suddenly from $h_0$ to another value, say $h$. The
initial Hamiltonian with $h_0$ for $t<0$ is denoted by ${\cal H}_0$, and its
ground state is $\left|\mathit{\Psi}^{(0)}_0\right\rangle$.  For $t>0$ the new
Hamiltonian ${\cal H}$ with $h$ governs the coherent time-evolution of the
system; for example an observable, represented by the operator $\hat{A}$, has
the time-evolution in the Heisenberg picture as: $\hat{A}(t)=\exp(\imath t
{\cal H})\hat{A}\exp(-\imath t {\cal H})$, and its expectation value for $t>0$
is given by $A(t)=\left\langle
\mathit\Psi^{(0)}_0\right|\hat{A}(t)\left|\mathit\Psi^{(0)}_0\right\rangle$.  
Dynamics of the system out of equilibrium is governed by the complete spectrum of ${\cal
H}$ and not only by the lowest excitations. Therefore, Hamiltonians with different
spectral properties will have completely different nonequilibrium properties. 

The form of the inhomogeneity in the couplings is generally crucial to the
spectrum of a Hamiltonian. For example the spectrum of the homogeneous quantum
Ising chain is absolutely continuous, thus all the eigenstates are extended.
By contrast, the random chain has a singular point spectrum and the eigenstates
are localized.  The spectrum of quasiperiodic chains lies between the above
mentioned two limiting cases \cite{suto,Damanik}; for example, the spectrum of
the Fibonacci chain defined in (\ref{hamilton}) is given by a Cantor set of
zero Lebesgue measure, signaling that the spectrum is of a multifractal type,
and it is called purely singular continuous \cite{Yessen} in the mathematical
denotation.  See \cite{suto} for precise mathematical
definitions of different spectra.

Below we first briefly review nonequilibrium properties
of the entanglement entropy and local magnetization
after a quench in the homogeneous chain and in random chains.

\subsection{Entanglement entropy}
\label{sec:entropy_hom}
The \textit{entanglement entropy}, ${\cal S}_{\ell}(t)$, of a block of the
first $\ell$ sites in the chain is defined as ${\cal S}_{\ell}(t)={\rm
Tr}_{\ell}[\rho_{\ell}(t) \ln\rho_{\ell}(t)]$ in terms of the reduced density
matrix: ${\bf \rho}_{\ell}(t)={\rm Tr}_{i>\ell}|\mathit{\Psi}_0(t)\rangle\langle
\mathit{\Psi}_0(t) |$. Here $|\mathit{\Psi}_0(t)\rangle$ denotes the ground state of the
complete system at time $t>0$. The details of the calculation of
$S_{\ell}\left(t\right)$ in the free-fermion representation can be found in the appendix
of \cite{ISzL09}.

For the \textit{homogeneous chain} (corresponding to the case with $r=1$
in (\ref{J_i})) in the limit $L \to \infty$ and for 
$\ell \gg 1$ the results can be summarized as follows \cite{Calabrese_05,Igloi_12}:
\be
S_{\ell}\left(t\right)=
\left\{ \begin{array}{ll}
\alpha t,\quad & t<\ell/v_{\mathrm{max}} \\
\beta \ell,\quad & t \gg \ell/v_{\mathrm{max}}\;,
\end{array} \right.
\label{S_t}
\ee
where $v_{\mathrm{max}}$ is a maximum  velocity.  For a quench to a quantum
critical point, the result in (\ref{S_t}) is a consequence of conformal
invariance \cite{Calabrese_05}; for other cases, this behavior can be explained
in the frame of a semiclassical (SC) theory \cite{Calabrese_05,Rieger_11}:
entanglement between the subsystem and its environment arises when two quantum
entangled quasiparticles, which are emitted at $t=0$ and move ballistically
with opposite velocities, arrive in the subsystem and in the environment
simultaneously.  The prefactors $\alpha=\alpha(h_0,h)$ and $\beta=\beta(h_0,h)$
have been exactly calculated \cite{Fagotti08} and these agree with the results
obtained from the SC theory \cite{Rieger_11}.  In \cite{eip09}
$\alpha(h_0=0,h)$ has been evaluated in a closed formula, which is a continuous
function of $h$ but at the critical point $h=1$, its second derivative
is logarithmically divergent.
 
In the \textit{random chain} the excitations are localized and therefore the
dynamical entanglement entropy approaches a finite limiting value.  When the
quench is performed to the random quantum critical point, the average entropy
increases ultra-slowly as $\log[\log(t)]$ \cite{Igloi_12}. This behavior can be
explained in terms of the strong disorder renormalization
group \cite{fisher,im,Igloi_12,Vosk_12}.

\subsection{Local magnetization}
\label{sec:magn_hom}

Another quantity we consider is the \textit{local magnetization}, $m_l(t)$, at
a position, $l$, of an open chain. Following Yang \cite{Yang_52} this is defined
for large $L$ as the off-diagonal matrix-element:
\be 
m_l\left(t\right)=\left\langle \mathit{\Psi}^{(0)}_0\left|\sigma_l^x(t)\right|\mathit{\Psi}^{(0)}_1\right\rangle\;,
\label{m_l_def}
\ee
where $\left|\mathit{\Psi}^{(0)}_1\right\rangle$ is the first excited state of the initial
Hamiltonian. Calculation of the magnetization in terms of free fermions 
is outlined in the appendix.

For the \textit{homogeneous chain} the time-dependence of the local
magnetization has been numerically calculated in
\cite{Rieger_11,Igloi_11}.  For the quench performed within the
ordered phases, $h_0<1$ and $h<1$, the results in the limit $L \to \infty$ and
$l \gg 1$ are given by:
\be
m_{l}\left(t\right) \sim
\left\{ \begin{array}{ll}
\exp(-t/\tau),\quad & t<l/v_{\mathrm{max}}\\
\exp(-l/\xi),\quad & t \gg l/v_{\mathrm{max}}\;,
\end{array} \right.
\label{m_l}
\ee
where the relaxation (decoherence) time $\tau$ and the correlation length $\xi$ depend on
the quench parameters $h_0$ and $h$. Exact expressions of these quantities have been 
derived recently \cite{Calabrese_11,Calabrese_12,Calabrese_12b}. In the small $h_0$ and $h$ limit,
accurate results can also been obtained from the SC theory \cite{Rieger_11,Sachdev_97}.
In this framework the quasiparticles in terms of the $\sigma$ operators are represented by ballistically 
moving kinks. Each time when a kink passes site $l$, the $\sigma_l^z$ operator changes sign;
thus kinks that pass a site an even number of times have no effect on the local magnetization.
Summing up contributions of all kinks, we obtain the functional form in (\ref{m_l}).
If the quench is performed close to the critical point, the kinks have a finite width;
this effect can be taken into account in a modified SC theory \cite{Rieger_11,blass},
which provides exact results. 

For quenches involving the disordered phase with $h_0>1$ and/or $h>1$, the
results obtained numerically \cite{Rieger_11,Igloi_11} or analytically
by the form-factor approach \cite{Calabrese_11,Calabrese_12,Calabrese_12b} indicate that
for bulk spins in large systems the first equation of (\ref{m_l}) is modified as:
\be
m_{l}\left(t\right) \simeq A(t)
\exp(-t/\tau)\;,
\label{m_losc}
\ee
where the prefactor, $A(t)$ changes sign during the relaxation process, say
$A(t)>0$ for $t_i < t <t_{i+1}$, $A(t)<0$ for $t_{i+1} < t <t_{i+2}$, etc.  The
period of these oscillations: $t_{\rm per}(h) \simeq (t_{i+1}-t_i)$ defines a
characteristic time-scale, which increases and becomes divergent as $h \to
1^+$. This is a signal of a \textit{dynamical phase transition} in the system.
The order parameter can be defined as:
\be
{\cal O}= \lim_{t \to \infty} \frac{1}{t}\int_0^t \left[ |A(t')| - A(t')\right] {\rm d} t'\;,
\label{order}
\ee
which is positive (${\cal O}>0$) in the oscillatory phase and ${\cal O}=0$ in the non-oscillatory phase.

In a \textit{disordered chain} away from the random quantum critical point
the bulk magnetization approaches a finite limiting value, which reflects
the localized nature of the excitations. After a quench performed to
the critical point, the average bulk magnetization has been found to
vanish asymptotically in a very slow way \cite{Igloi_unp},
$m_\mathrm{b}(t) \sim [\ln (t)]^{-A}$, where $A>0$ is a disorder dependent constant.

\section{Results for quasiperiodic chains}
\label{sec:quasi}
In this section we present our results for the quasiperiodic quantum Ising
chain after a global quench, obtained by numerical calculations based on the
free-fermion representation of the model. We concentrate on the Fibonacci chain
with the parameter $\omega$ defined in (\ref{f_i}) being the golden mean.
We consider finite chains with a length fixed at a Fibonacci number $F_n$.  We have
calculated the entanglement entropy and the local magnetization for system
sizes up to $L=F_{17}=1597$. For the numerical calculation we solved hermitien
and anti-hermitien eigenvalue problems, and calculated complex determinants using
the LAPACK routine. For a given set of parameters ($h_0,h$ and $r$) the time-dependence
of the entropy or the magnetization of a chain with $L=1597$ 
was obtained in about one day of CPU time on a $2.5$ GHz processor.

Below we present results for these two quantities
separately.

\subsection{Entanglement entropy}
\label{sec:entropy}


\begin{figure}[h!]
\begin{center}
\includegraphics[width=6cm, clip]{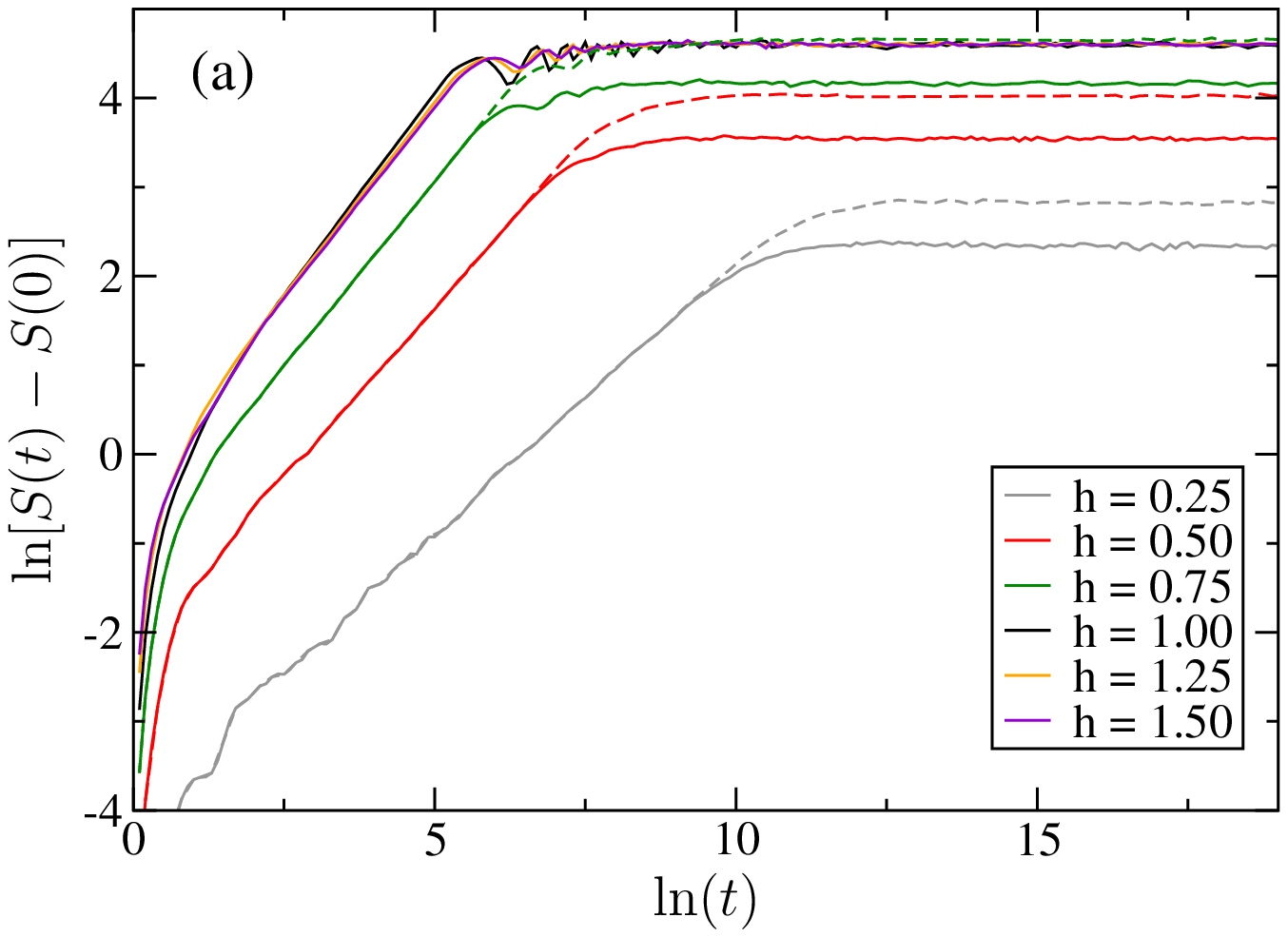}
\includegraphics[width=6cm, clip]{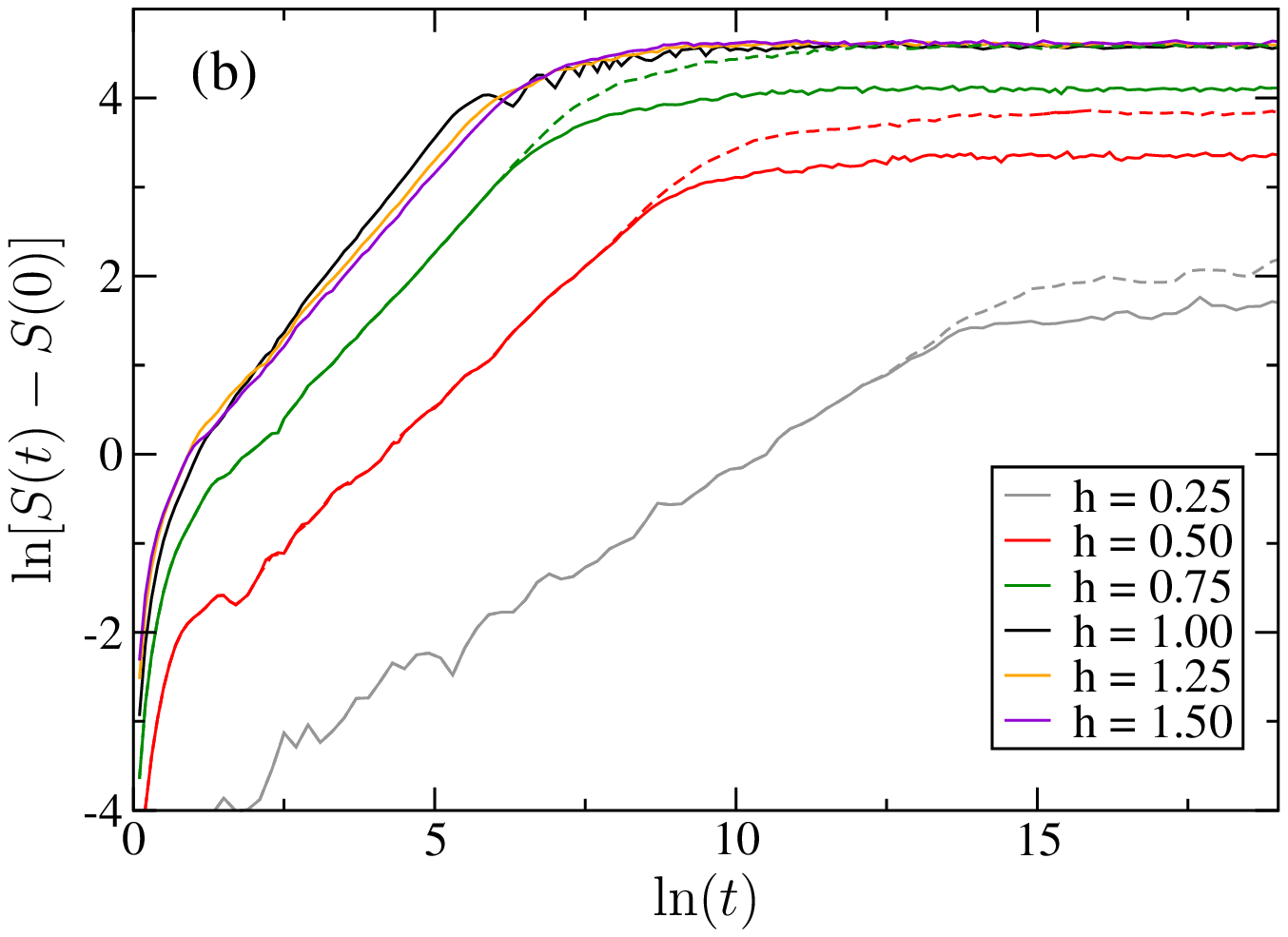}
\includegraphics[width=6cm, clip]{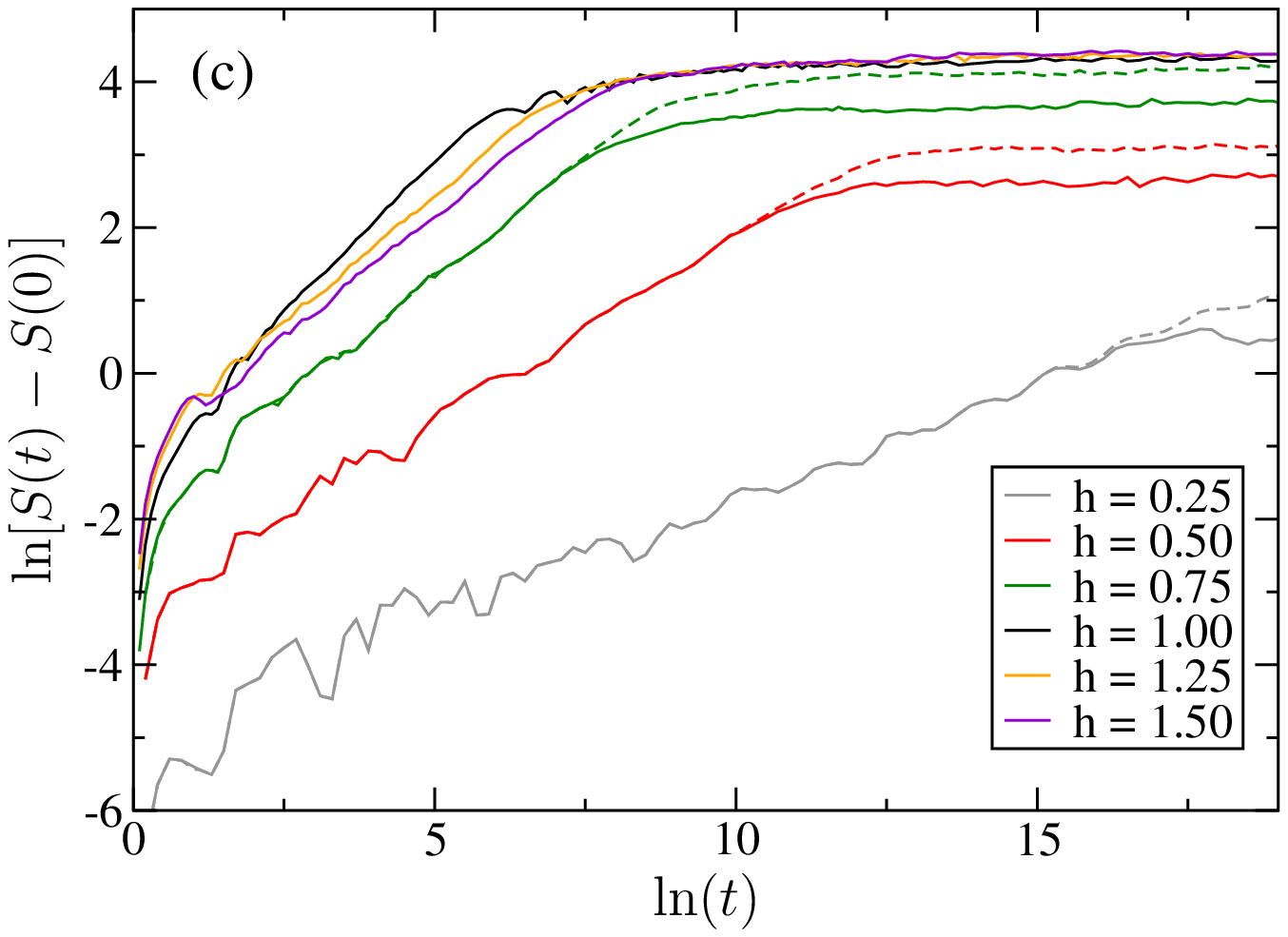}
\end{center}
\caption{Dynamical entropy after a quench from $h_0=0$ to various values of $h$ at the aperiodicity parameters 
(a) $r=0.75$, (b) $r=0.5$ and (c) $r=0.25$. The solid lines are the results for $L=F_{16}=987$, and 
the dashed lines (only at $h=0.25$, $h=0.5$ and $h=0.75$) correspond to the data for $L=F_{17}=1597$.
The ''noise'' (irregular variation) present on the curves in the small $t$ regime is due to such
low-energy excitations, which are related to local properties of the quasiperiodic chain and are 
independent of the chain lengths.}
\label{fig1}
\end{figure}


For a chain of total length $F_n$ with
periodic boundary conditions, we have calculated the entanglement entropy
$S_\ell$ between a block of length $\ell=F_{n-2}$ and its environment which
has a length of $F_{n-1}$. Various values of $0<r<1$ for the inhomogeneity
amplitude were considered.  We start our numerical calculations from the fully
ordered state with $h_0=0$ to a state with $h>0$ both in the ordered and 
in the disordered phases, as well as at the critical point. The numerical results
for ${\cal S}_{\ell}(t)-{\cal S}_{\ell}(0)$ are shown in figure~\ref{fig1}.
For all cases considered, $S_\ell(t)$ exhibits two time-regimes: 
in the late-time regime, the entropy is saturated to 
an $L$ dependent value, similar to the behavior for the homogeneous chain;
in the early-time regime, it increases with time as a power-law form:
\be
{\cal S}(t) \sim t^{\sigma}\;,
\label{sigma}
\ee
with some exponent $\sigma<1$. Our numerical results show that the exponent 
$\sigma$ depends on the value of the transverse field in the final state, while it does not vary (significantly)
with the initial $h_0$. The values of $\sigma$ for $r=0.25$, $0.5$ and $0.75$
are plotted in figure~\ref{fig5}; for all cases considered, $\sigma$ reaches
its maximum at the critical point $h=1$, and the increase with $h$ in
the ordered phase ($h<1$) is much faster than the decrease in the
disordered phase ($h>1$). Furthermore, we have found that
the exponent $\sigma$ decreases with stronger inhomogeneity, that is
with smaller value of $r$.


\begin{figure}[h]
\begin{center}
\includegraphics[width=8cm, angle=0, clip]{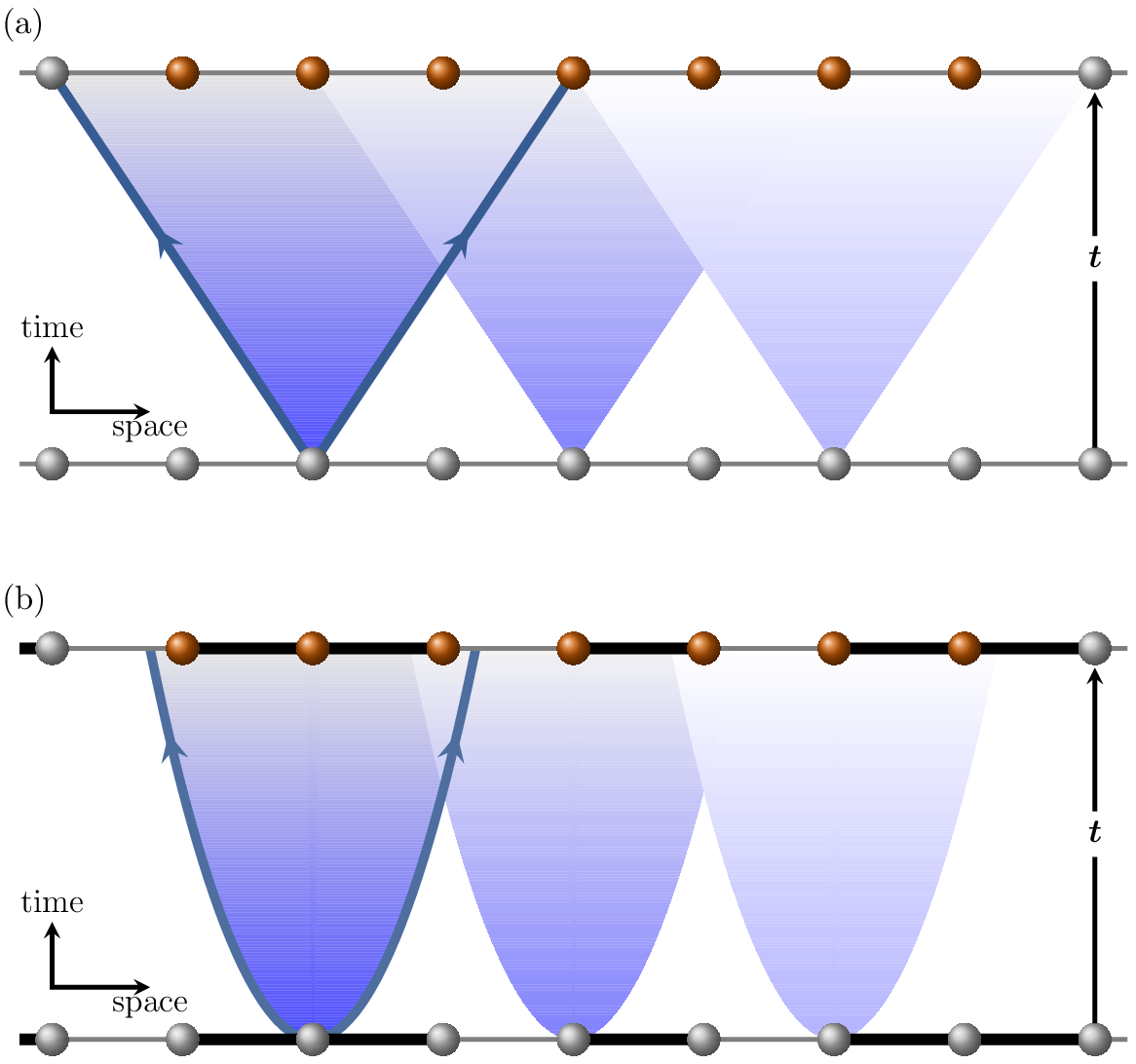}
\end{center}
\caption{Schematic illustration of the light cones of quasiparticles
for a homogeneous quantum Ising chain (a) and for a chain with an aperiodic modulation 
of the couplings (the thin/thick lines between sites represent weak/strong couplings according
to a Fibonacci sequence) (b). The quasiparticle excitations emitted at time $t=0$  move ballistically in 
the homogeneous lattice, 
while their motion is anomalous diffusive with $x \sim t^D$ ($D<1$) in the quasiperiodic lattice. 
Pairs of quasiparticles moving to the left or right from a given point are entangled;
they will contribute to the entanglement entropy between a region $A$ (the region with
orange sites) and the rest of the chain, region $B$, if they arrive simultaneously in $A$ and $B$.}
\label{fig1A}
\end{figure}


The power-law time-dependence of the entanglement entropy in (\ref{sigma}) is a
new feature of the quasiperiodic system: the increase in entropy is slower than
in the homogeneous chain, but faster than in a random chain.  This behavior can
be explained in terms of quasiparticles that are emitted at time $t=0$, and
subsequently move classically by anomalous diffusion which has a power-law
relationship between displacement and time, $x \sim t^D$, with a diffusion
exponent $0<D<1$.  We note that in a homogeneous chain pairs of quasiparticles
that contribute to the entanglement entropy move ballistically (\textit{i.e.}
$x\sim t$) rather than moving by diffusion, which results in the linear growth
of the entanglement entropy with time \cite{Calabrese_05} (figure 2).  The dynamics of the
quasiparticles in our quasiperiodic lattice will be studied in more detail in
section~\ref{sec:wave}.

\subsection{Local magnetization}
\label{sec:magn}


\begin{figure}[h]
\begin{center}
\includegraphics[width=6cm, angle=0, clip]{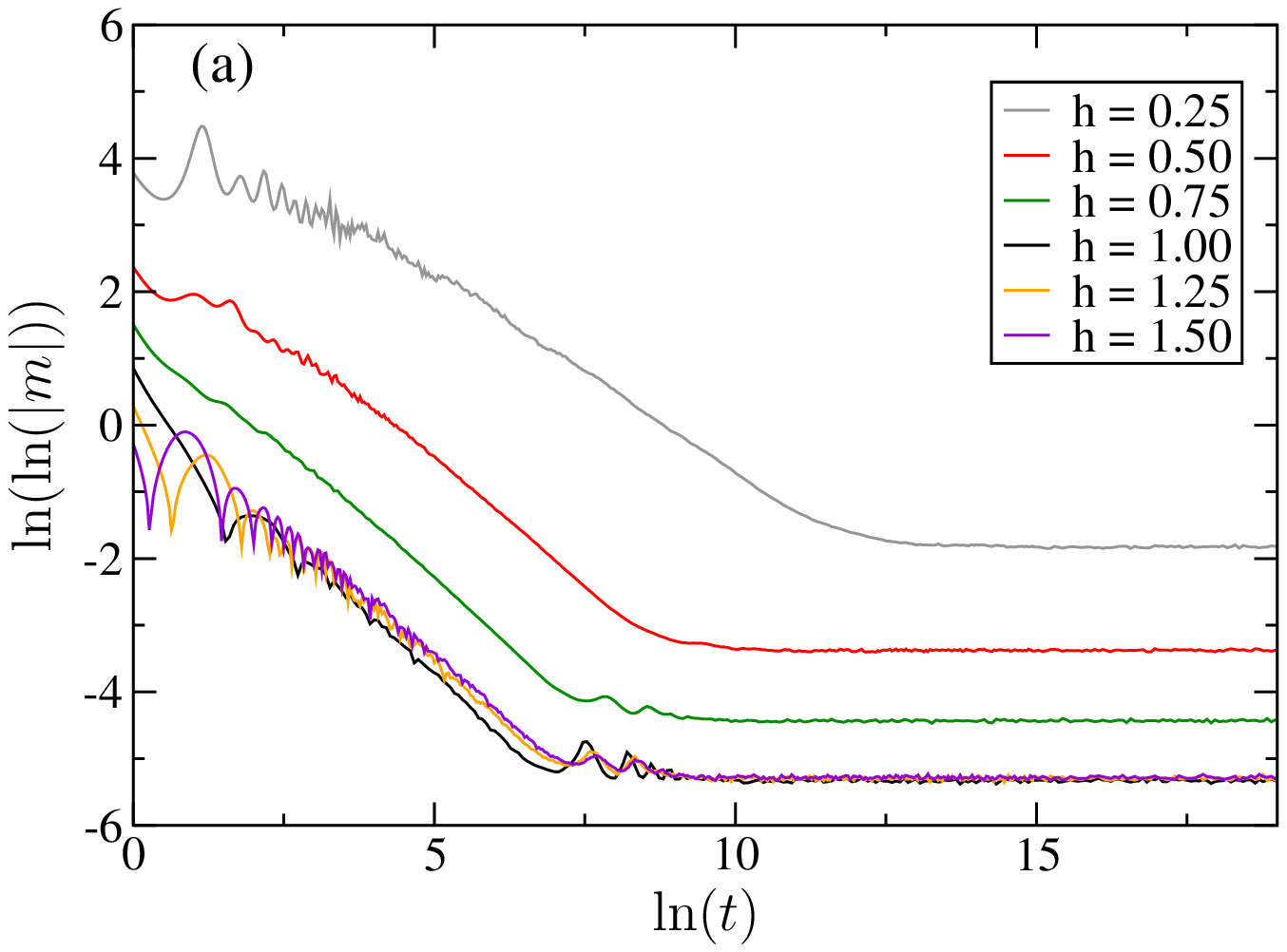}
\includegraphics[width=6cm, angle=0, clip]{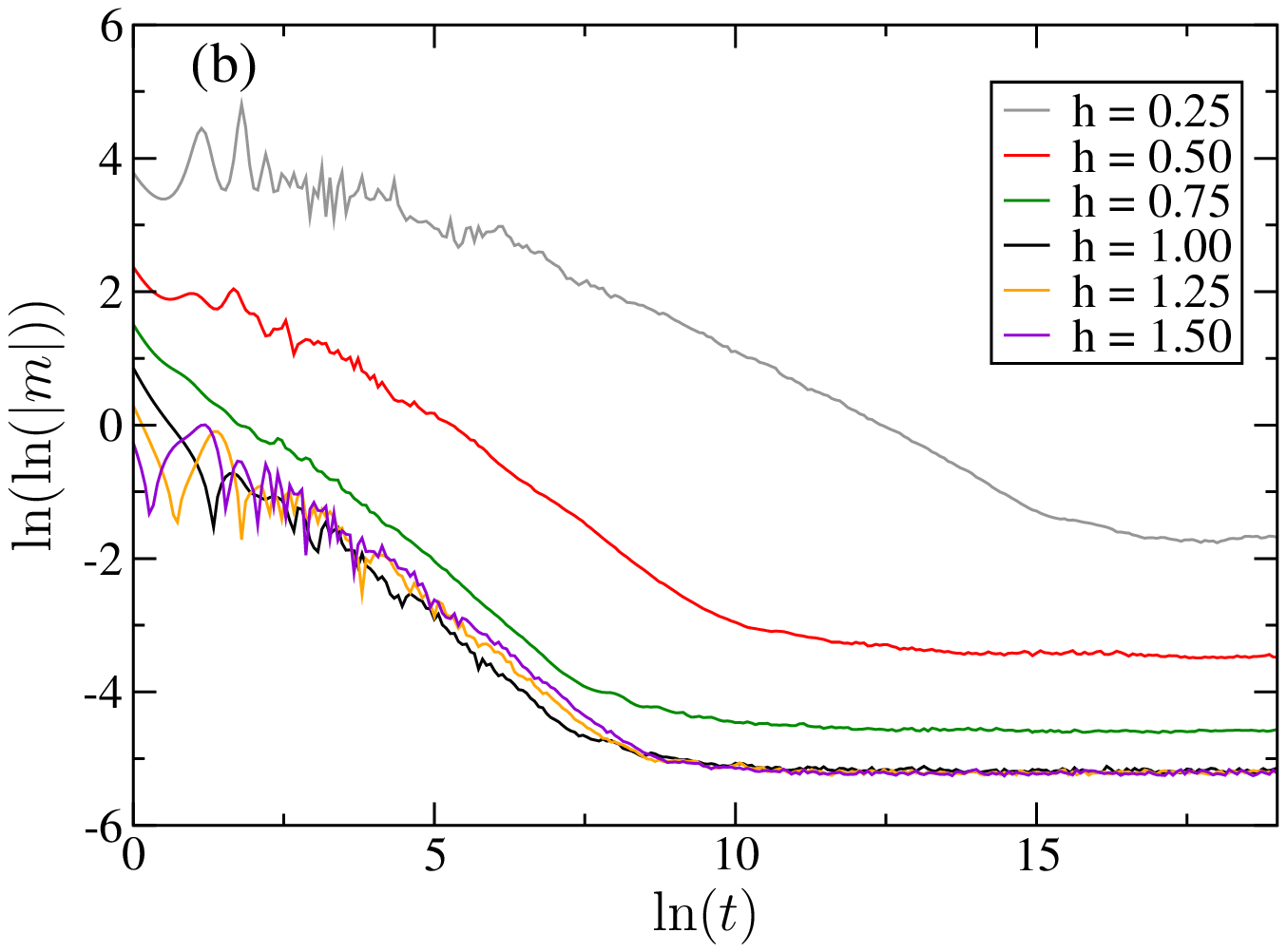}
\includegraphics[width=6cm, angle=0, clip]{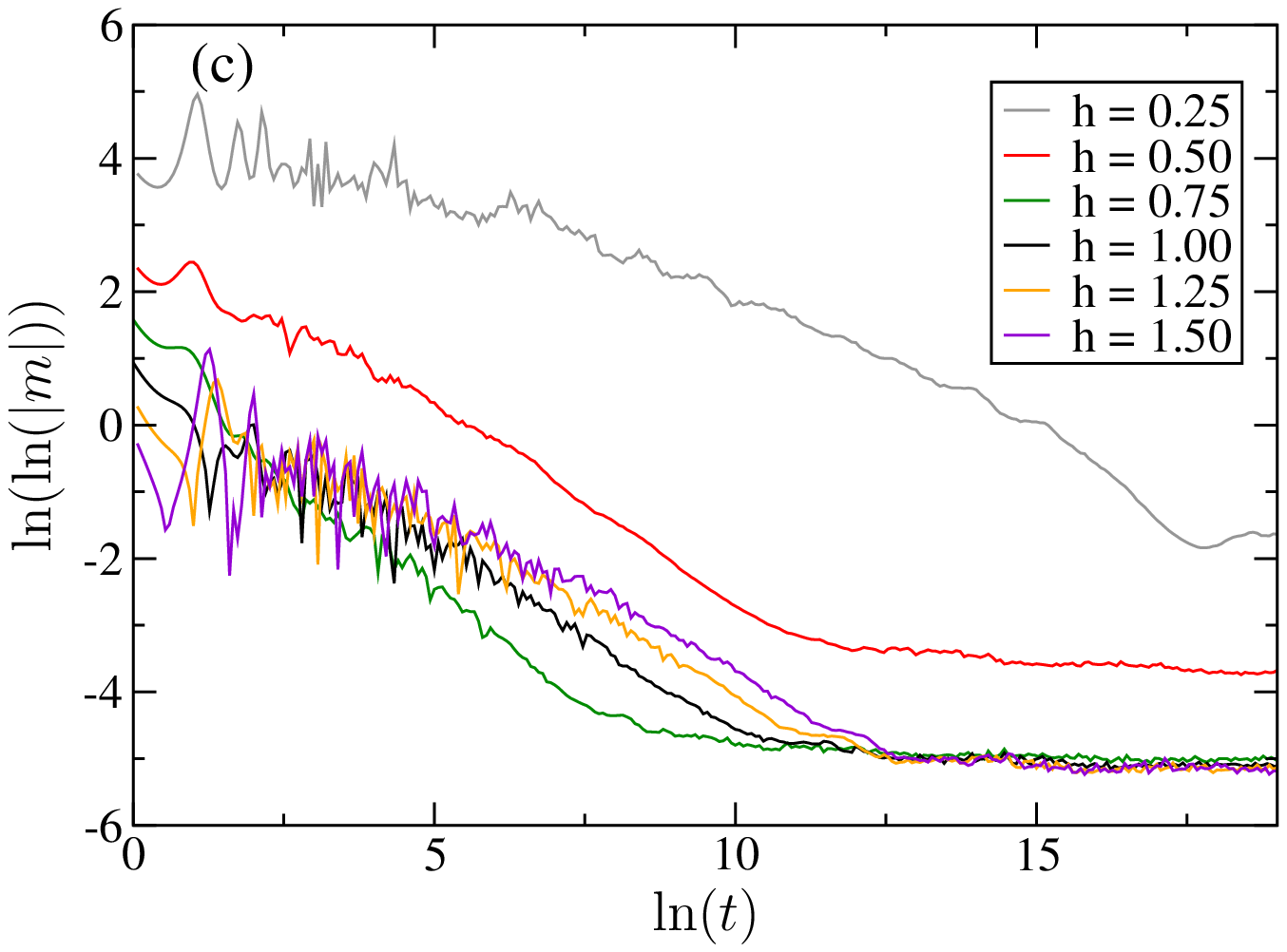}
\includegraphics[width=6cm, angle=0, clip]{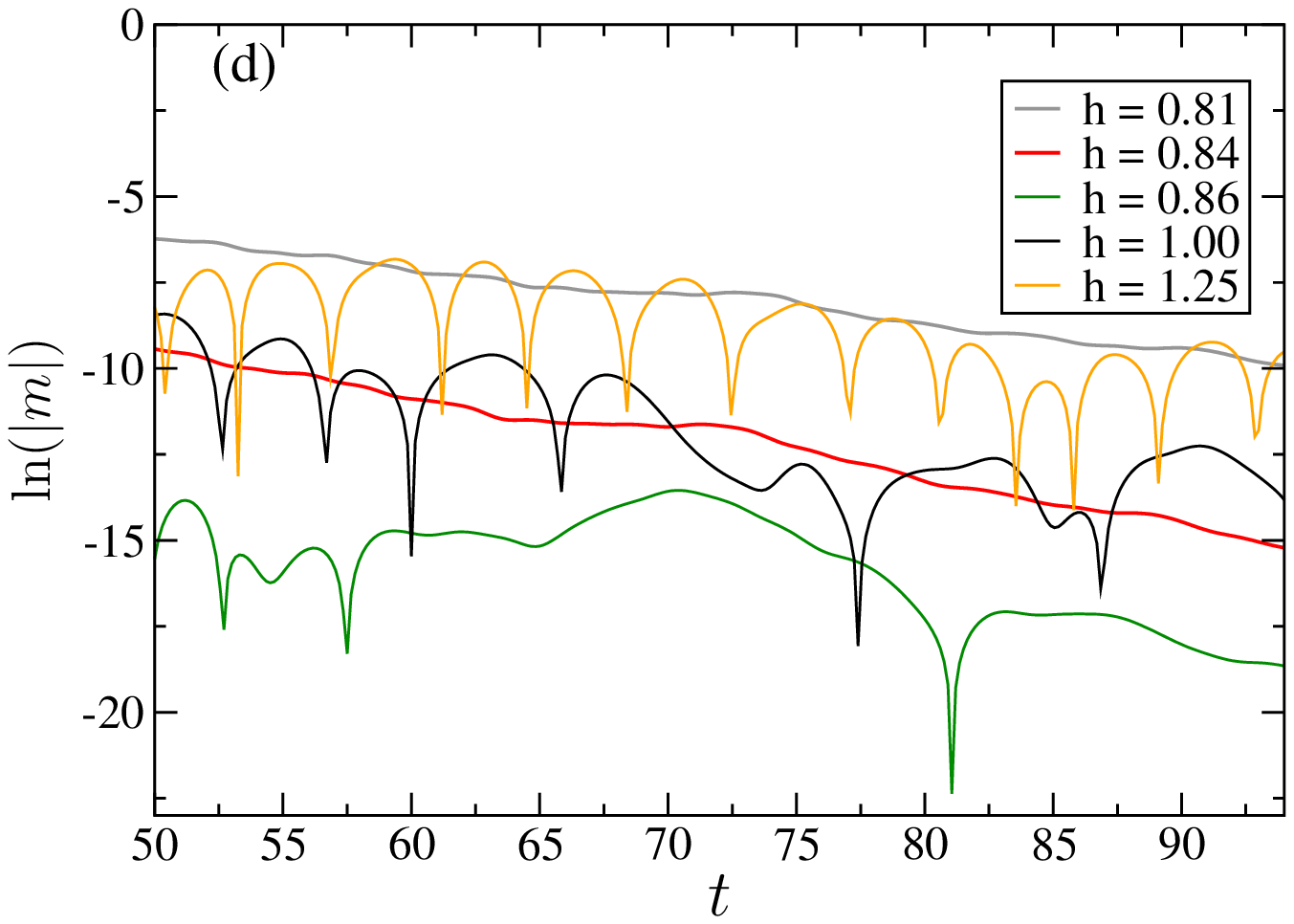}
\caption{
Double logarithm of the bulk magnetization as a function of the logarithm of
the time. During the quench the transverse field is changed from $h_0=0$ to
different values of $h$ at the aperiodicity parameter $r=0.75$ (panel (a)),
$r=0.5$ (panel (b)), $r=0.25$ (panel (c)). The length of the chain is
$L=F_{17}=1597$ and the magnetization is considered at site $l=F_{16}=987$. 
In panel (d) $\ln|m_\mathrm{b}(t)|$ is shown as a function of $t$ 
in the window $50<t<100$ for different values of $h$ at $r=0.5$. The oscillations
in $\ln|m_\mathrm{b}(t)|$ ({\it i.e.} in the prefactor $A(t)$)
occur when $h$ is larger than a certain value $h^*$ (here $h^*\approx 0.85$),
and the oscillations disappear for $h<h^*$; the dynamical phase transition
described in the main text occurs at $h^*$.
}
\label{fig2}
\end{center}
\end{figure}


The local magnetization, $m_l(t)$, is calculated for open chains of length
$L=F_n$. Generally $m_l(t)$ has a monotonic position dependence:
$m_{l_1}(t)>m_{l_2}(t)$ for $l_1 < l_2<L/2 $.  We measured the magnetization at
site $l=F_{n-1}$, which is considered as the bulk magnetization and
denoted by $m_\mathrm{b}(t)$. We have also studied the behavior of the surface
magnetization, $m_1(t)$, for which some exact results are obtained.

We study the asymptotic behavior of the surface magnetization
(given in (\ref{m_1})) for large $t$ after a quench.  If the quench is performed to the
ordered phase, $h<1$, the lowest excitation energy is $\epsilon_1=0$
({\it i.e.} $\cos(\epsilon_1 t)=1$); consequently $P_{1,2k-1}(t)$ in (\ref{P}) has a time
independent part. This results in a non-oscillating contribution
to the surface magnetization: 
$\overline{m_1}=\lim_{t\to\infty} \int_0^t m_1(t'){\rm d} t'$
which is given by:
\be
\overline{m_1}=\Phi_1(1) \sum_{j=1}^L \Phi_1(j)\Phi_1^{(0)}(j)\;,
\label{m_1_asymp}
\ee
and defines its stationary value. Recall that $\Phi_1(1)=m_1(h,t=0)$, i.e. it is equal to the equilibrium
surface magnetization \cite{peschel84,igloi_rieger98}, which is finite  for $h<1$, 
and zero in the disordered phase. Similarly, $\Phi_1^{(0)}(1)>0$ for $h_0<1$ and zero otherwise. 
From this it follows that the stationary nonequilibrium surface magnetization is $\overline{m_1}>0$, if both $h<1$ and $h_0<1$. 
Otherwise the stationary surface magnetization vanishes.
If the quench starts from the fully ordered initial state $h_0=0$, then $\Phi^{(0)}_1(j)=\delta_{1,j}$ and
$\overline{m_1}=\Phi_1^2(1)$; thus we obtain the simple relation:
\be
\overline{m_1}(h)=[m_1(h,t=0)]^2\;,
\label{m_1_rel}
\ee
which is generally valid between the stationary value of the nonequilibrium
surface magnetization and its equilibrium value. From (\ref{m_1_rel}) it follows that the
critical exponent $\beta_\mathrm{s}^{\mathrm{ne}}$ for the nonequilibrium surface magnetization and
the critical exponent $\beta_\mathrm{s}$ for the equilibrium surface magnetization are related as: 
$\beta_\mathrm{s}^{\mathrm{ne}}=2 \beta_\mathrm{s}$.  
According to (\ref{m_1_rel}) and \cite{irt99}, 
for the Fibonacci chain close to the critical point $h\to h_c=1$,
we have $\overline{m_1}(h) \sim 1-h^2=(h_c-h)(h_c+h)\sim h_c-h$, thus $\beta_s^{\mathrm{ne}}=1$.


\begin{figure}[!th]
\begin{center}
\includegraphics[width=8cm, angle=0, clip]{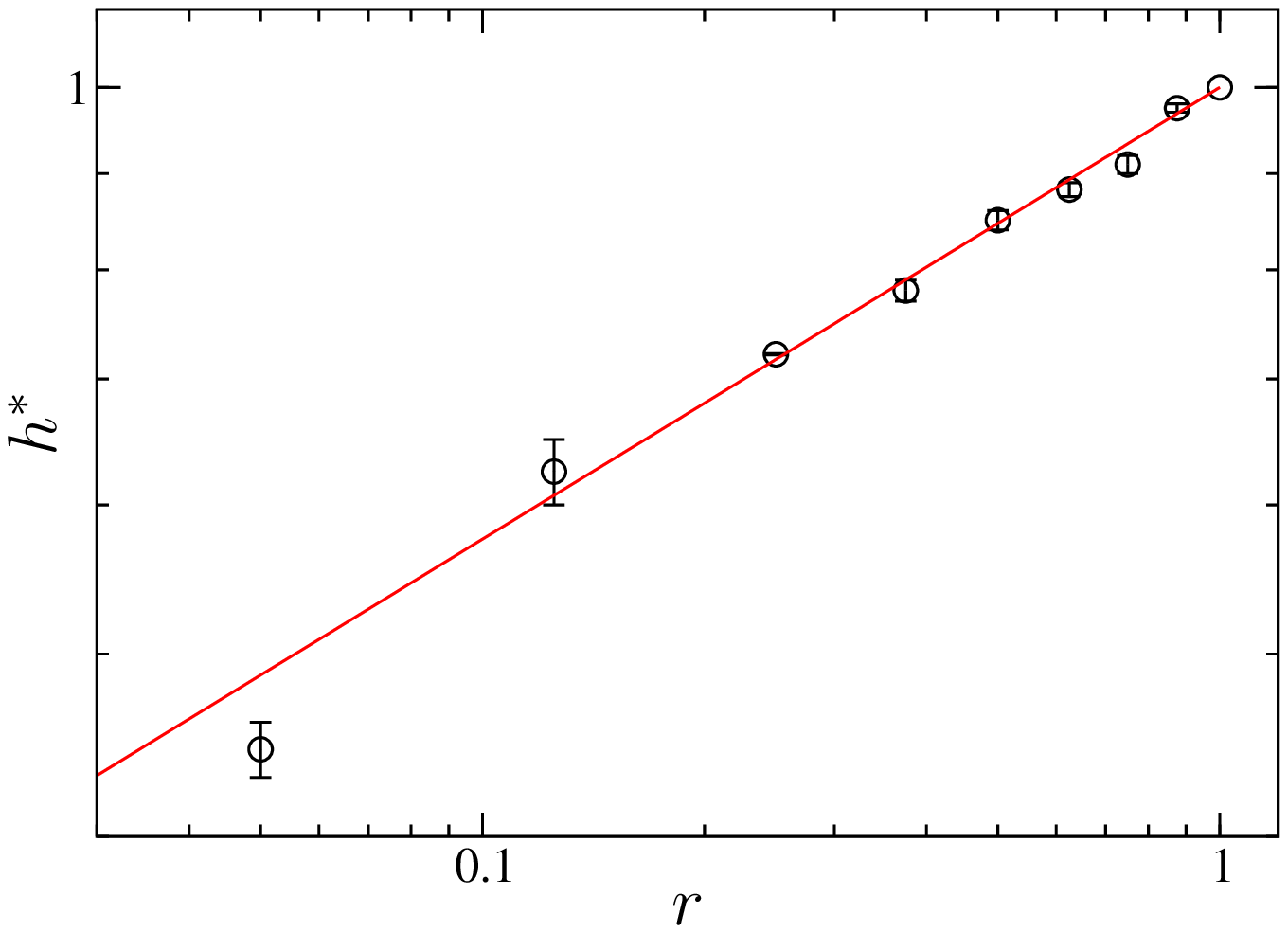}
\caption{Position of the dynamical critical point for different values of the aperiodicity
parameter in a double-logarithmic plot. The straight line has a slope $\alpha=0.24$.}
\label{fig3}
\end{center}
\end{figure}


We numerically calculated the time-dependence of the bulk magnetization after a
quench from the fully ordered initial state, $h_0=0$, to different values of
$h$. For fixed values of the inhomogeneity $r=0.25,\,0.5,\,0.75$, the results
for the double logarithm of $|m_\mathrm{b}(t)|$ are shown in figure~\ref{fig2}(a-c) as
functions of $\ln t$. In each case one can observe a linear dependence, which
implies that the magnetization has asymptotically a stretched exponential time
dependence:
\be
m_\mathrm{b}(t) \sim A(t) \exp\left(-C t^{\mu}\right)\;.
\label{mu}
\ee
which corresponds to equation (\ref{m_losc}) for a homogeneous
system, with $\mu=1$. Before analyzing the decay exponent $\mu$, we first
study the behavior of the prefactor $A(t)$. Like in a homogeneous chain as
discussed in section \ref{sec:magn_hom}, there is a dynamical phase transition
between a non-oscillating phase for $h<h^*(r)$, where the order-parameter
${\cal O}$ defined in (\ref{order}) is zero, and an oscillating phase for
$h>h^*(r)$, where ${\cal O}>0$. In the oscillating phase, the characteristic
time-scale defined as the period time, $t_{\rm per}(h,r)$, becomes divergent as
$h \to h^*(r)^+$. An example for this behavior is illustrated in
figure~\ref{fig2}, panel (d), in which $\ln|m_\mathrm{b}(t)|$ as a function of
$t$ is shown in the window $50<t<100$ for different values of $h$ at $r=0.5$;
as seen in this figure, the curves for $h=0.86, 1.0$ and $1.25$ oscillate,
whereas the oscillations vanish for $h=0.81$ and $h=0.84$. We identify the 
dynamical phase transition point as $h^*=0.850(5)$. In this quasiperiodic model the
dynamical phase transition does not coincide with the equilibrium phase
transition, since $h^*(r)<1$ for $r<1$.  Estimates of $h^*(r)$ versus $r$ are
shown in figure \ref{fig3}; the data are well approximated by a power-law
$h^*(r) \sim r^{\alpha}$ with $\alpha=0.24(3)$ \cite{h*}.

The exponent $\mu$ describing the decay of the local magnetization dependents 
both on $h$ and $r$; by contrast, it does not vary significantly with $h_0$, at least for $h_0 < h$. 
Our results for the critical exponents $\mu=\mu(h,r)$ are plotted in figure~\ref{fig5} for $r=0.75,~0.5$ and
$0.25$ as functions of $h$. The exponent $\mu$ reaches its maximum at the dynamical phase transition point
$h^*(r)$.

\subsection{Interpretation by wave packet dynamics}
\label{sec:wave}
As is known from previous studies on the homogeneous chain, dynamical features of
the entanglement entropy and the local magnetization can be well described by
the dynamics of quasiparticles. To understand the dynamical properties of the
quasiparticles emitted after a quantum quench in the quasiperiodic lattice, we
regard the quasiparticles as wave packets and study their dynamics using a
method that has been applied to studies of transport properties of
quasicrystals \cite{Mayou,wave_packet}.


\begin{figure}
\begin{center}
\includegraphics[width=6cm, angle=0, clip]{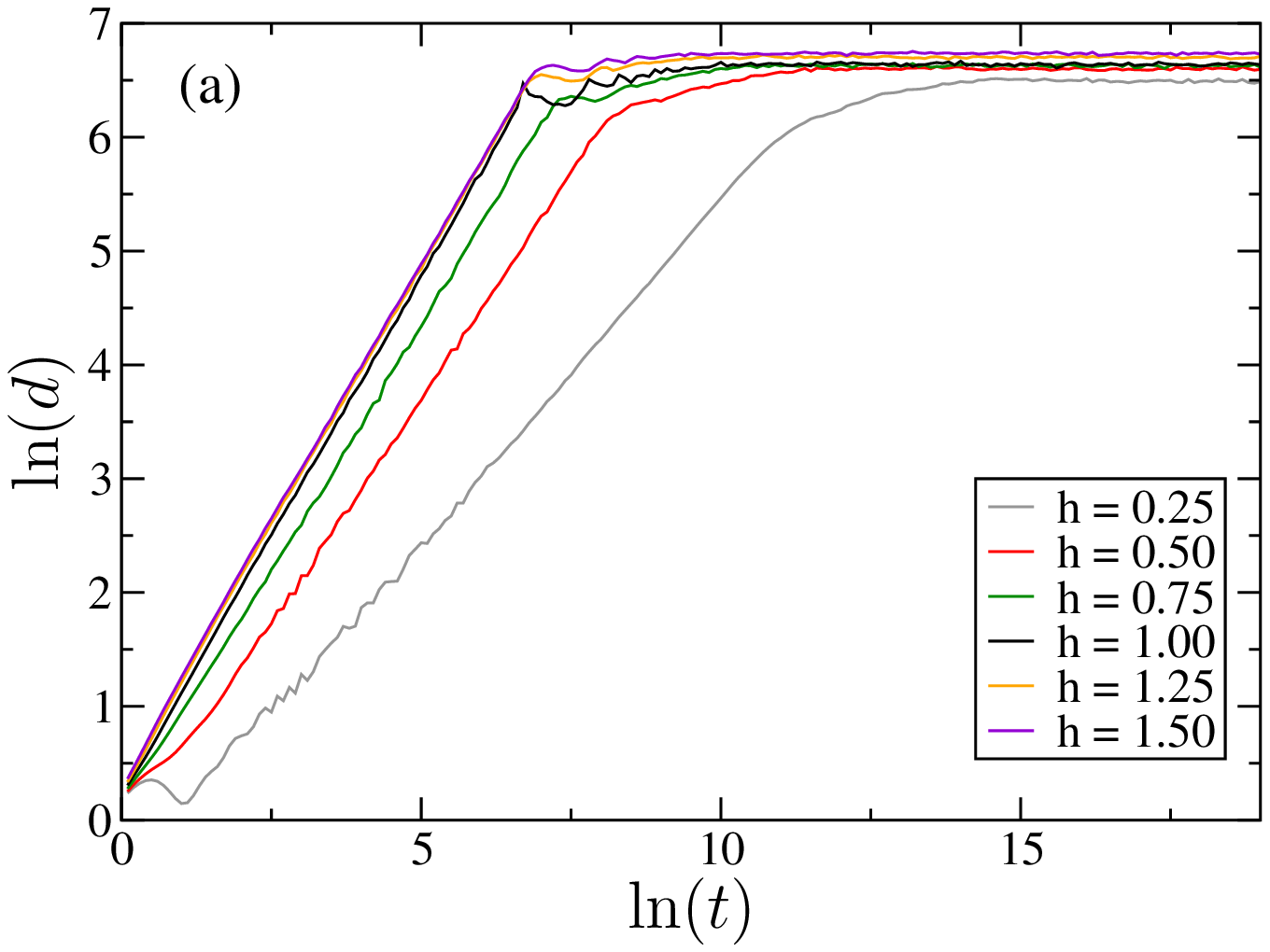}
\includegraphics[width=6cm, angle=0, clip]{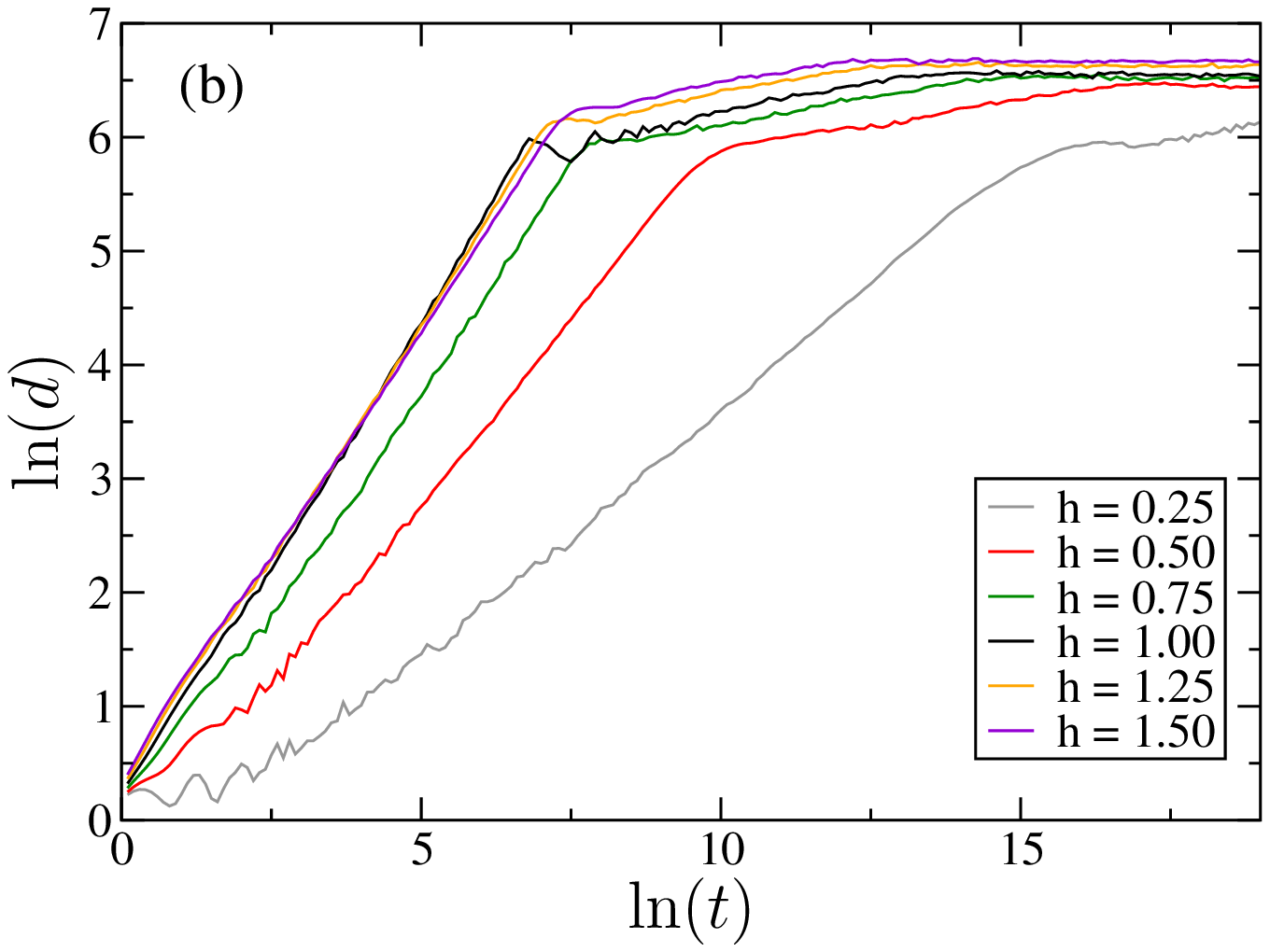}
\includegraphics[width=6cm, angle=0, clip]{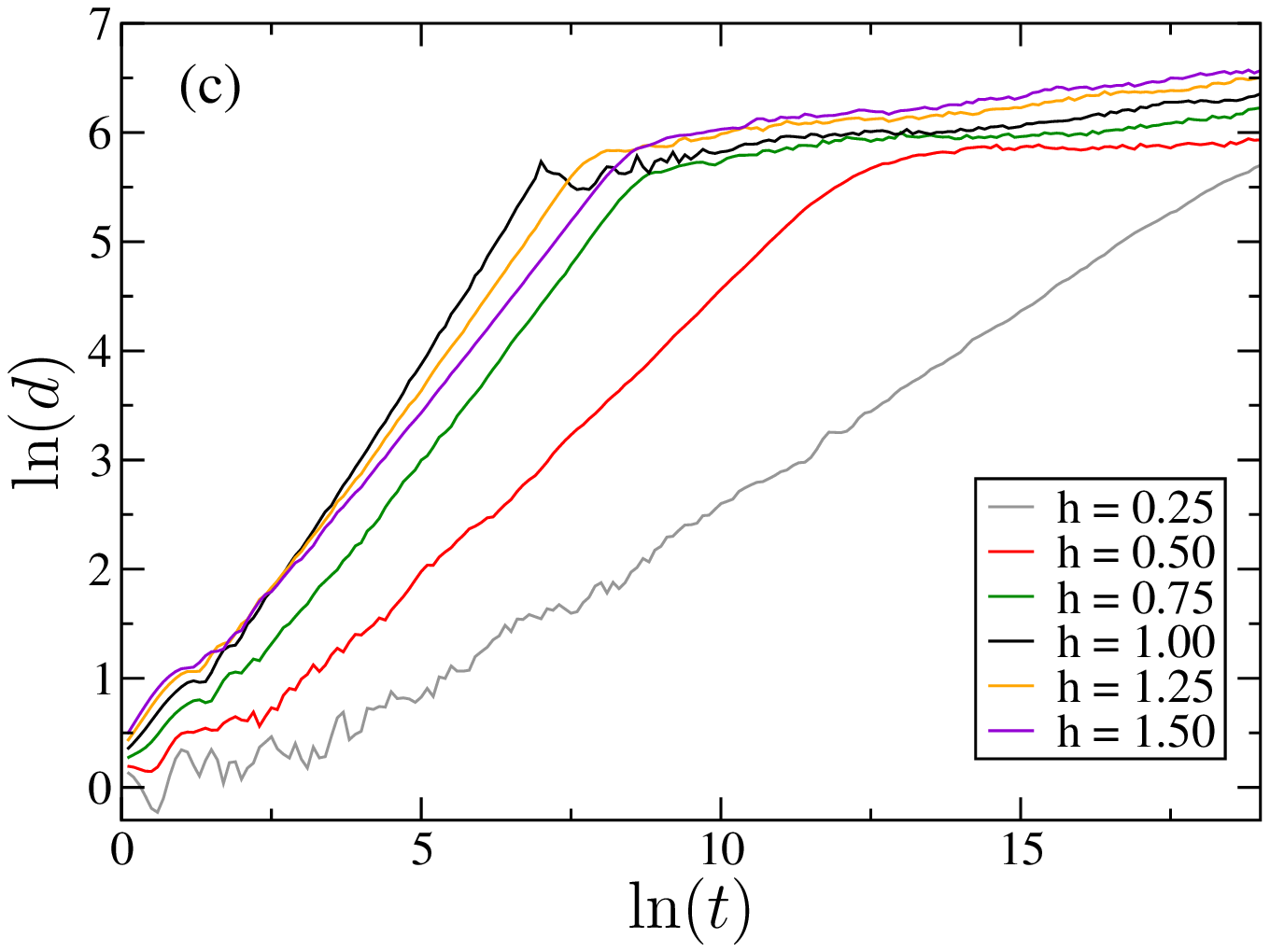}
\end{center}
\caption{Time-dependent width of the wave-packet at different values
of $h$ for $r=0.75$ (panel (a)), $r=0.5$ (panel (b)), $r=0.25$ (panel (c)).
}
\label{fig4}

\end{figure}


We construct a wave packet connecting sites $k$ and $l$ at time $t$ in the form:
\beqn
 W_{l,k}(t) =&\frac{1}{2}&\sum_q \biggl\{ \cos( \epsilon_q t) \bigl[\Phi_q(l) \Phi_q(k)+\Psi_q(l) \Psi_q(k) \bigr]\biggr.\nonumber \\ 
           &-& \biggl.\imath \sin( \epsilon_q t) \bigl[\Phi_q(l) \Psi_q(k)+ \Phi_q(k) \Psi_q(l) \bigr] \biggr\},
 \label{W}
\eeqn
which is localized at $t=0$ since $W_{l,k}(0)=\delta_{l,k}$ ({\it cf.} equation (\ref{normalization})). 
For a Hamiltonian with eigenfunctions $\phi_q(l)$ and
eigenvalues $\epsilon_q$, a wave packet can be obtained by: $W_{l,k}(t)=\sum_q  \cos( \epsilon_q t) \phi_q(l) \phi_q(k)$,           
which corresponds to the first term in (\ref{W}).
We note that (\ref{W}) is just
a linear combination of the four time-dependent factors in (\ref{P}), which describe the
time dependence of the fermion operators. 
The width of the wave packet starting from site $k$ after time $t$ is given by:
\be
d(k,t)=\left[ \sum_l (k-l)^2 |W_{l,k}(t)|^2\right]^{\frac{1}{2}}\;.
\ee
The spreading of a wave packet in a perfect crystal with absolutely continuous
energy spectrum is known to be ballistic, {\it i.e.} the width increases
linearly in time. A heuristic argument is the following \cite{Thouless}: the
energy scale $\Delta \epsilon$  defined by the typical variation of the energy
levels is proportional to the inverse of the time that a wave packet needs to
spread over the chain.   In the case of the absolutely continuous spectrum, we
have $\Delta \epsilon \sim L^{-1}$, which gives $d\sim L\sim t$. In case of a
singular continuous spectrum as for our quasiperiodic lattice, there are many
energy scales $\Delta \epsilon \sim L^{-1/\alpha}$ with a number of exponents
$\alpha$.  One then expects that for large $t$ the wave packet in the infinite
quasiperiodic lattice shows anomalous diffusion in the form $d(k,t) \sim
t^{D(k)}$ with a diffusion exponent $D(k)$, which may depend on the starting
position.  Here we determine the value of $D(k)$ numerically.

After a global quench, quasiparticles are emitted everywhere in lattices,
therefore $d(k,t)$ should be averaged over different initial positions,
\be
d(t)=\overline{d(k,t)} \sim t^D\;.
\label{d(t)}
\ee
In our numerical calculations chains of length $L=F_{17}=1597$ with periodic boundary conditions
were considered.  First we have confirmed that the wave packet constructed in our
method moves ballistically in the homogeneous chain (with $r=1$), corresponding to
$D=1$.  In the quasiperiodic chains the motion is indeed anomalous diffusive
with $D<1$, which is seen in figure~\ref{fig4} where the average widths of the
wave packet are presented as functions of time in a log-log plot for various
values of $h$ and $r=0.75,~0.5$ and $0.25$. The diffusion exponent $D$
for given $h$ and $r$ corresponds to the slope of the linear part of the function.

\begin{figure}
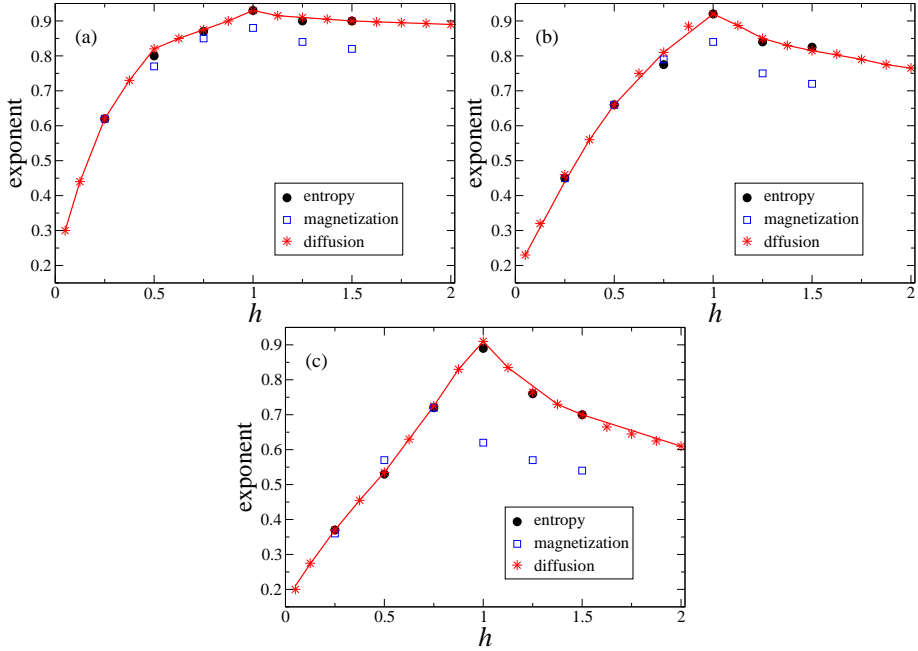

\begin{center}
\includegraphics[width=6cm, clip]{fig6a.eps}
\includegraphics[width=6cm, clip]{fig6b.eps}
\includegraphics[width=6cm, clip]{fig6c.eps}
\end{center}
\caption{Scaling exponents calculated from the time-dependence
of the width of the wave-packet, from the entanglement entropy and from the
magnetization at different values
of $h$ for $r=0.75$ (panel (a)), $r=0.5$ (panel (b)), $r=0.25$ (panel (c)).
The full lines connecting the diffusion exponents are guides for the eye.}
\label{fig5}

\end{figure}

The variation of $D$ with $h$ at a fixed $r$ is shown in figure~\ref{fig5}, 
compared with the exponent $\sigma$ for the entanglement entropy and 
the exponent $\mu$ for the local magnetization. Here one can observe that the 
agreement between these three exponents is very good for
$h<h^*(r)$, i.e. in the non-oscillating phase, but the exponent for the magnetization deviates in the oscillating 
phase ($h>h^*(r)$).
The discrepancy in the oscillating phase implies that the semiclassical picture breaks down
in the oscillating phase, where the quasiparticles cannot be well described by
the moving kinks in the magnetization.


\section{Discussion}
\label{sec:disc}

In this paper we have studied the nonequilibrium dynamics of quasiperiodic
quantum Ising chains after a global quench. In a quench process, the complete
spectrum of the Hamiltonian is relevant for the the time evolution of various
observables.  For the quasiperiodic quantum Ising chain the spectrum is in a
very special form, which is given by a Cantor set of zero Lebesgue measure,
{i.e.} purely singular continuous. We have calculated numerically two
quantities: the dynamical entanglement entropy and the relaxation of the local
magnetization. The entanglement entropy is found to increase in time as a
power-law (see (\ref{sigma})), whereas the bulk magnetization decays in a
stretched exponential way (see (\ref{mu})). Both behaviors can be explained in
a quasiparticle picture, in which the quasiparticles move by anomalous
diffusion in the quasiperiodic lattice. The diffusion exponent has been
calculated by a wave packet approach, and good agreement has been found with
the exponents that we obtained for the entropy and for the magnetization. We
note that the anomalous dynamics found in the global quench process is similar
to the transport properties of quasicrystals.

Relaxation of the bulk magnetization is found to present a nonequilibrium
dynamical phase transition.  The non-oscillating phase, in which the
magnetization is always positive, and the oscillating phase, in which the sign
of the magnetization varies periodically in time, is separated by a dynamical
phase transition point, at which the time-scale of oscillations diverges. This
singularity point, due to collective dynamical effects, is different from the
equilibrium critical point.

A similar nonequilibrium dynamical behavior is expected to hold for other
quasiperiodic or aperiodic quantum models as long as the spectrum of the
Hamiltonian is also purely singular continuous; there is a large class of such
models, for example the Thue-Morse quantum Ising chain. If, however the
spectrum of the Hamiltonian of the model is in a different type, such as the
Harper potential which has extended or localized states, the nonequilibrium dynamics 
is expected to be different than the case we consider in this paper.

\ack{FI is grateful to D. Karevski, H. Rieger and A. S\"ut\H o for discussions. FI acknowledges support from
the Hungarian National Research Fund under grant No OTKA K75324 and K77629; he
also acknowledges travel support from the National Science Council (NSC),
Taiwan, under Grant No.~101-2912-I-004-518. YCL is supported by the NSC under
Grants No. NSC~98-2112-M-004-002-MY3, No. NSC~101-2112-M-004-005-MY3 and
No.~NSC~100-2923-M-004-002-MY3; she also gratefully acknowledges support from the National Center for Theoretical
Sciences of Taiwan.} 
\vskip -.5cm

\appendix

\section{Free-fermionic calculation of the time-dependent local magnetization}
\label{sec:app}

To calculate the local magnetization in (\ref{m_l_def}),
we need to first calculate the time dependence of the spin operator
$\sigma_l^x(t)$ at site $l$ in the Heisenberg picture.
We introduce at each site two Majorana fermion
operators, $\check{a}_{2l-1}$ and $\check{a}_{2l}$, defined in terms of
the free fermion operators $\eta_k^\dag$ and $\eta_k$ (given in (\ref{eta_ferm})) as
\beqn
\check{a}_{2l-1}&=&\sum_{k=1}^L \Phi_k(l)(\eta_k^\dag +\eta_k)\; ,\nonumber\\
\check{a}_{2l}&=&-\imath \sum_{k=1}^L \Psi_k(l)(\eta_k^\dag -\eta_k) \;.
\label{majorana}
\eeqn
These satisfy the commutation relations:
\be
\check{a}_{l}^{\dag}=\check{a}_{l},\quad \{ \check{a}_{l},\check{a}_{k} \}=2\delta_{l,k}\;.
\label{comm_maj}
\ee
The spin operators are then expressed in terms of the Majorana operators as:
\begin{eqnarray}
\sigma_l^x&=&\;\imath^{l-1} \prod_{j=1}^{2l-1} \check{a}_{j} \;,
\label{sigma_xz}
\end{eqnarray}
and the local magnetization in (\ref{m_l_def})
is then given as the expectation value of product of fermion operators
with respect to the ground state:
\be
m_l(t)=(\imath)^{l-1}\langle \mathit{\Psi}^{(0)}_0|\prod_{j=1}^{2l-1} \check{a}_{j}(t) \eta_1 |\mathit{\Psi}^{(0)}_0 \rangle\;,
\label{Wick}
\ee
where we have used: $|\mathit{\Psi}^{(0)}_1 \rangle=\eta_1 |\mathit{\Psi}^{(0)}_0 \rangle$.
The expression in (\ref{Wick}) - according to Wick's theorem - can be expressed as a sum of products of two-operator
expectation values. This can be written in
a compact form of a Pfaffian, which in turn can be evaluated as the square root of the determinant of an
antisymmetric matrix:

%
\beqn
\fl
m_l(t)
= (-\imath)^{l-1}\left\vert
\begin{array}{cccc}
\langle \check{a}_1(t)\check{a}_2(t)\rangle &
\langle \check{a}_1(t)\check{a}_3(t)\rangle &
\cdots\quad
\langle \check{a}_1(t)\check{a}_{2l-1}(t_1)\rangle &
\langle \check{a}_1(t)\eta_1\rangle \\
&
\langle \check{a}_2(t)\check{a}_3(t) \rangle &
\cdots\quad
\langle \check{a}_2(t)\check{a}_{2l-1}(t)\rangle &
\langle \check{a}_2(t)\eta_1\rangle  \\
 &          \ddots & & \vdots\cr
  &         &  \langle \check{a}_{2l-2}(t)\check{a}_{2l-1}(t)\rangle &
\langle \check{a}_{2l-2}(t) \eta_1\rangle \\
  &          & &  \langle \check{a}_{2l-1}(t) \eta_1\rangle
\end{array}
\right\vert\nonumber\\
= \pm \left[ {\rm det}\, C_{ij}\right]^{1/2}\;,
\label{pfaffian}
\eeqn
%
where $C_{ij}$ is the antisymmetric matrix $C_{ij}=-C_{ji}$, with the
elements of the Pfaffian (\ref{pfaffian}) above the diagonal. (Here and in the following
we use the short-hand notation: $\langle \dots \rangle =\langle \mathit{\Psi}^{(0)}_0 |\dots
| \mathit{\Psi}^{(0)}_0 \rangle$.)

Below we describe how the time evolution of the spin operator $\sigma_l^x$ follows from
the time dependence of the Majorana fermion operators. Inserting
$\eta_k^\dag(t)=e^{\imath t\epsilon_k} \eta_k^\dag$ and $\eta_k(t)=e^{-\imath t\epsilon_k} \eta_k$ into (\ref{majorana}) one
obtains
\be
\check{a}_{m}(t)=\sum_{n=1}^{2L} P_{m,n}(t) \check{a}_{n}\;,
\label{a_t}
\ee
with
\beqn
P_{2l-1,2k-1}&=&\sum_q \cos( \epsilon_q t) \Phi_q(l) \Phi_q(k) ,\nonumber\\
P_{2l-1,2k}&=&-\sum_q \sin (\epsilon_q t) \Phi_q(l) \Psi_q(k)\; ,\nonumber\\
P_{2l,2k-1}&=&\sum_q \sin (\epsilon_q t) \Phi_q(k) \Psi_q(l)\; ,\nonumber\\
P_{2l,2k}&=&\sum_q \cos( \epsilon_q t) \Psi_q(l) \Psi_q(k)\;.
\label{P}
\eeqn
The two-operator expectation values are given by:
\beqn
\langle \check{a}_{m}(t)\check{a}_{n}(t)\rangle=\sum_{k_1,k_2} P_{m,k_1}(t)P_{n,k_2}(t)\langle\check{a}_{k_1}\check{a}_{k_2}\rangle\;.
\label{a_a}
\eeqn
The \textit{equilibrium} correlations in
the initial state with a transverse field $h_0$ are:
\beqn
\langle \check{a}_{2m-1}\check{a}_{2n-1} \rangle=\langle \check{a}_{2m}\check{a}_{2n} \rangle&=&\delta_{m,n},\cr
\langle \check{a}_{2m-1}\check{a}_{2n} \rangle=-\langle \check{a}_{2m}\check{a}_{2n-1} \rangle&=&\imath G^{(0)}_{n,m}\;,
\eeqn
where the static correlation matrix $G^{(0)}_{m,n}$ is given by:
\be
G^{(0)}_{m,n}=-\sum_q \Psi_q^{(0)}(m) \Phi_q^{(0)}(n)\;,
\ee
where $\Psi_q^{(0)}(m)$ and $\Phi_q^{(0)}(n)$ are the components of the eigenvectors 
in (\ref{Phi_Psi}), calculated for the initial Hamiltonian.
Then (\ref{a_a}) can be written in the form:
\be
\langle \check{a}_{m}(t)\check{a}_{n}(t) \rangle = \delta_{m,n} + \imath \Gamma_{m,n}(t)\;,
\label{corr_maj}
\ee
with
%
\beqn
\Gamma_{2l-1,2m-1}&=&\sum_{k_1,k_2} \left[ G^{(0)}_{k_2,k_1} P_{2l-1,2k_1-1} P_{2m-1,2k_2}\right. \cr
&-&\left. G^{(0)}_{k_1,k_2} P_{2l-1,2k_1} P_{2m-1,2k_2-1} \right] \cr
\Gamma_{2l-1,2m}&=&\sum_{k_1,k_2} \left[ G^{(0)}_{k_2,k_1} P_{2l-1,2k_1-1} P_{2m,2k_2}\right. \cr
&-&\left. G^{(0)}_{k_1,k_2} P_{2l-1,2k_1} P_{2m,2k_2-1} \right] \cr
\Gamma_{2l,2m-1}&=& -\sum_{k_1,k_2} \left[ G^{(0)}_{k_2,k_1} P_{2l,2k_2} P_{2m-1,2k_1-1}\right. \cr
&-&\left. G^{(0)}_{k_1,k_2} P_{2l,2k_2-1} P_{2m-1,2k_1} \right] \cr
\Gamma_{2l,2m}&=&\sum_{k_1,k_2} \left[ G^{(0)}_{k_2,k_1} P_{2l,2k_1-1} P_{2m,2k_2}\right. \cr
&-&\left. G^{(0)}_{k_1,k_2} P_{2l,2k_1} P_{2m,2k_2-1} \right]\,.
\eeqn
In (\ref{pfaffian}) there are also the contractions:
\beqn
\Pi_m&=&
\langle \mathit{\Psi}^{(0)}_0 |\check{a}_{m}(t)\eta_1|\mathit{\Psi}^{(0)}_0 \rangle \cr
&=&\sum_n P_{m,n} \langle \mathit{\Psi}^{(0)}_0 |\check{a}_{n}\eta_1|\mathit{\Psi}^{(0)}_0 \rangle
\label{Pi}
\eeqn
where
\beqn
\langle \mathit{\Psi}^{(0)}_0 |\check{a}_{2l-1}\eta_1|\mathit{\Psi}^{(0)}_0 \rangle&=& \Phi_1^{(0)}(l) \cr
\langle \mathit{\Psi}^{(0)}_0 |\check{a}_{2l}\eta_1|\mathit{\Psi}^{(0)}_0 \rangle&=& \imath \Psi_1^{(0)}(l)\;.
\eeqn
Thus finally the square of the local magnetization is given by the determinant:
\beqn
m_l^2(t)
=
\left\vert
\begin{array}{cccccc}
0 &
\Gamma_{1,2} &
\Gamma_{1,3} &
\cdots\quad
\Gamma_{1,2l-1} &
\Pi_{1} \cr
-\Gamma_{1,2}&
0 &
\Gamma_{2,3} &
\cdots\quad
\Gamma_{2,2l-1} &
\Pi_{2}\cr
-\Gamma_{1,3}&
-\Gamma_{2,3}&
0 &
\cdots\quad
\Gamma_{3,2l-1} &
\Pi_{3}\cr
 & &         & \ddots &  \vdots\cr
-\Gamma_{1,2l-1} &  \cdots\quad        &  & 0 & \Pi_{2l-1}\cr
-\Pi_1&  \cdots\quad&          &   -\Pi_{2l-1} &0 
\end{array}
\right\vert
\label{pfaffian1}
\eeqn
%
As a special case, the surface magnetization is expressed as:
\beqn
m_1(t)=\Pi_1&=&\sum_{j=1}^L P_{1,2j-1}(t) \Phi_1^{(0)}(j) \cr
&-&\imath\sum_{j=1}^L P_{1,2j}(t) \Psi_1^{(0)}(j) \;.
\label{m_1}
\eeqn


\begin{thebibliography}{99}
\bibitem{h*}
No oscillation of the magnetization is expected if all sites are ``locally''
in the ferromagnetic phase. This condition is satisfied for a weakly coupled site having
one strong ($J_s$) and one weak ($J_w$) bond, and if $\ln h<\ln J_s + \ln J_w$, which means
$h<r^{2/\omega-1}$. The numerical results in figure~\ref{fig3} indicate that the critical value, $h^*(r)$
coincides with $r^{2/\omega-1}$.
\end{thebibliography}
\end{document}